\documentclass[aps,prapplied,reprint,groupedaddress]{revtex4-2}

\usepackage{amsmath,amssymb}
\usepackage{graphicx}
\usepackage{bm} 
\usepackage{hyperref}

\newcommand{\e}{{\rm e}}

\usepackage[dvipsnames]{xcolor}

\usepackage{color}

\graphicspath{{./}}

\begin{document}
\title{Analysis of High Performing Terahertz Quantum Cascade Lasers}
\author{Viktor Rindert}
\affiliation{Mathematical Physics and
  NanoLund, Lund University, Box 118, 22100 Lund, Sweden}
\author{Ekin  {\"O}nder}
\affiliation{Mathematical Physics and NanoLund, Lund
  University, Box 118, 22100 Lund, Sweden}
\author{Andreas Wacker}
\email{Andreas.Wacker@teorfys.lu.se}
\affiliation{Mathematical Physics
  and NanoLund, Lund University, Box 118, 22100 Lund, Sweden}
\date{September 30, 2022; accepted as Letter in Physical Review Applied}
\begin{abstract}
Detailed simulations for Terahertz quantum cascade lasers based on
two-well designs  are presented. We reproduce the maximal operation
temperatures observed and attribute the degradation with temperature
to the occupation of parasitic levels and thermal
back-filling. Furthermore,  we demonstrate that the current injection
can be conveniently studied by using EZ states, which combine energy selectivity and spatial localization. Improving the
injection allows to achieve higher maximal operation temperatures
around 265 Kelvin. 
\end{abstract}
\maketitle
The quantum cascade laser (QCL) \cite{FaistScience1994} is one of the
most relevant devices for IR radiation, in particular since room
temperature operation could be achieved \cite{BeckScience2002}. This
is still a main goal for QCLs operating in the Terahertz (THz)
region \cite{KohlerNature2002} (i.e. below the optical phonon
frequency). While there had been no improvement for a long time after
scratching  the 200 K mark in 2012 \cite{FathololoumiOE2012}, recently
a significant improvement was achieved based on two-well designs
resulting in thermoelectrically cooled \cite{BoscoAPL2019} and
portable \cite{KhalatpourNatPhot2021} systems operating up to 250 K.
More information on the different design ideas and their benefits can
be found in the recent review \cite{WenProgQuantumElectron2021}.

In this work we focus on two-well designs, where essentially three
levels are of relevance. 
Next to the upper (u) and lower (l) laser level, the ground (g)
level extracts electrons from the lower level by phonon emission
and injects them into the upper level
of the subsequent module by resonant tunneling. This is the minimal
design for any QCL \cite{KumarAPL2009B,ScalariOE2010,WackerAPL2010}
and has thus the benefit of a rather simple and robust design allowing
for short modules with high gain per length.  Upon increasing the
doping density and barrier height following detailed simulations,
these structures became superior to other designs
\cite{FranckieAPL2018,BoscoAPL2019}. Using even higher barriers, the
current record operational temperature of 250 K was reported in 2021
\cite{KhalatpourNatPhot2021}.

In Ref.~\cite{KhalatpourNatPhot2021} it was argued that the key
benefit of their design is the establishment of a pure 3-level
design, where other levels are carefully separated in energy and
space. This also reduces leakage, an important issue for devices with
a short module length
\cite{KhanalJOpt2014,AlboAPL2015B,AlboAPL2017,WangSciRep2021,MiyoshiJAP2021}.
In this article we want to investigate this point  for the
structures with highest operation temperature. We show that the
increased energy separation between the lower laser and ground level
plays an important role, which reduces the thermal back-filling as
already suggested in
\cite{BaranovAIPAdvances2019,FranckiePhysRevApplied2020,DemicNJP2022}.
Furthermore, we
analyze the injection in detail. Based on these considerations,
we propose a structure, named LU2022, which according to our
modeling operates at even  higher temperatures.

The calculations presented here are based on our non-equilibrium
Green's function (NEGF) code, detailed in
\cite{WackerIEEEJSelTopQuant2013}. Unless mentioned otherwise all
parameters used in this work are given in Ref.~\cite{WingeJAP2016},
where the versatility and accuracy of the model is detailed. We use
seven levels per module\footnote{This includes all Wannier levels
    below $\Delta E_c$ and at least one level above $\Delta E_c$ for
    all samples, so that all possible tunneling events among bound
    states are safely included.}  and repeat the modules 2 times in
each direction in the simulations in order to allow for tunneling over
more than one module boundary and the population of higher parasitic
states (indexed by p).  We note that the temperature used in the
calculations mainly\footnote{It also enters the screening of the
  potential  from ionized impurities.} determines the thermal
occupation of the phonon modes and is referred to as phonon
temperature $T_\textrm{ph}$ in the following.

We assume total losses of 20/cm for the metal-metal waveguides used in
\cite{BoscoAPL2019,KhalatpourNatPhot2021} around 4 THz. This is
consistent with data from \cite{HanOptExpress2018} and was also used
in \cite{WingeJAP2016}.

It was noted earlier \cite{WingeJAP2016}, that our NEGF code
systematically produced higher currents than observed for samples
reported from the MIT group, when using our standard conduction band
offset  $\Delta E^1_c=x\times0.831\textrm{ eV}$ \cite{YiPRB2010}. In
contrast much better agreement in currents was found for samples from
the ETH group. In this respect it is interesting to note, that both
groups internally use different band offsets, when they show
band-diagrams. The ETH group \cite{BoscoAPL2019} uses values that are
consistent with $\Delta E^1_c$. In contrast, a higher value is used by
the MIT group. From the figures of \cite{KhalatpourNatPhot2021} and
information given in the supplementary information, we extracted
$\Delta E^2_c=x\times 1.01\textrm{ eV}$ based on the
Al$_x$Ga$_{1-x}$As bandgap reported in \cite{WasilewskiJAP1997}
together with a phenomenological 72\% share of the conduction band
offset. 

\begin{table}[b]
\caption{Experimental and simulation results using different band
  offsets $\Delta E_c^{1/2}$ for the devices ETH2019 from
  \cite{BoscoAPL2019} and MITG552/652 from
  \cite{KhalatpourNatPhot2021}. The simulations where performed for
  $T_\textrm{ph}=100$ K and the experimental values where extracted
  for the lowest temperature displayed.}
\label{table:ResultOffset}   
  \centering
\begin{tabular}{|l|c|c|c|}
\hline Wafer  & ETH2019 & MITG552 & MITG652\\ \hline
$J_{\textrm{max}}^{\textrm{exp}} [\textrm{kA/cm}^2]$ & 3.5 & 2.67 &
2.6\\ $J_{\textrm{max}}^{\textrm{sim}}  [\textrm{kA/cm}^2]$ for
$\Delta E_c^1$ & 4.3 & 3.3  & 3.6 \\ $J_{\textrm{max}}^{\textrm{sim}}
[\textrm{kA/cm}^2]$ for $\Delta E_c^2$ & 2.8 & 2.5 & 2.7 \\ \hline
$J_{\textrm{th}}^{\textrm{exp}} [\textrm{kA/cm}^2]$ & 2 & 1.48 &
1.54\\ $J_{\textrm{th}}^{\textrm{sim}}  [\textrm{kA/cm}^2]$ for
$\Delta E_c^1$ & 2.6 & 1.9 & 2.4 \\ $J_{\textrm{th}}^{\textrm{sim}}
[\textrm{kA/cm}^2]$ for $\Delta E_c^2$ & 1.4 & 1.2 & 1.4 \\ \hline
\end{tabular}  
\end{table}

Table \ref{table:ResultOffset} provides simulation results for
threshold  and peak currents (under lasing conditions where gain
matches the losses) for the QCL devices of
\cite{BoscoAPL2019,KhalatpourNatPhot2021}. We find good quantitative
agreement if $\Delta E^2_c$ is applied for the MIT devices. On the
other hand, $\Delta E^1_c$ is slightly better (albeit possibly too
small)  for the ETH device. We do not have an explanation for this
difference, which is consistent with the observations in
\cite{WingeJAP2016}. A straightforward explanation could be that the
Al-content $x$ is calibrated differently in both labs. On the other
hand, the geometrical definition of barriers might be
different. Furthermore, a part of the difference, may be attributed to
nonlinear behavior of $\Delta E_c$ in $x$, as the ETH device as a
lower Al content.  In the following we use $\Delta E^1_c$ for the ETH
sample of \cite{BoscoAPL2019} and $\Delta E^2_c$ for the MIT samples
of \cite{KhalatpourNatPhot2021}, i.e. we use the values given by the
respective groups.

\begin{figure}  \centering
 \includegraphics[width=0.8\columnwidth]{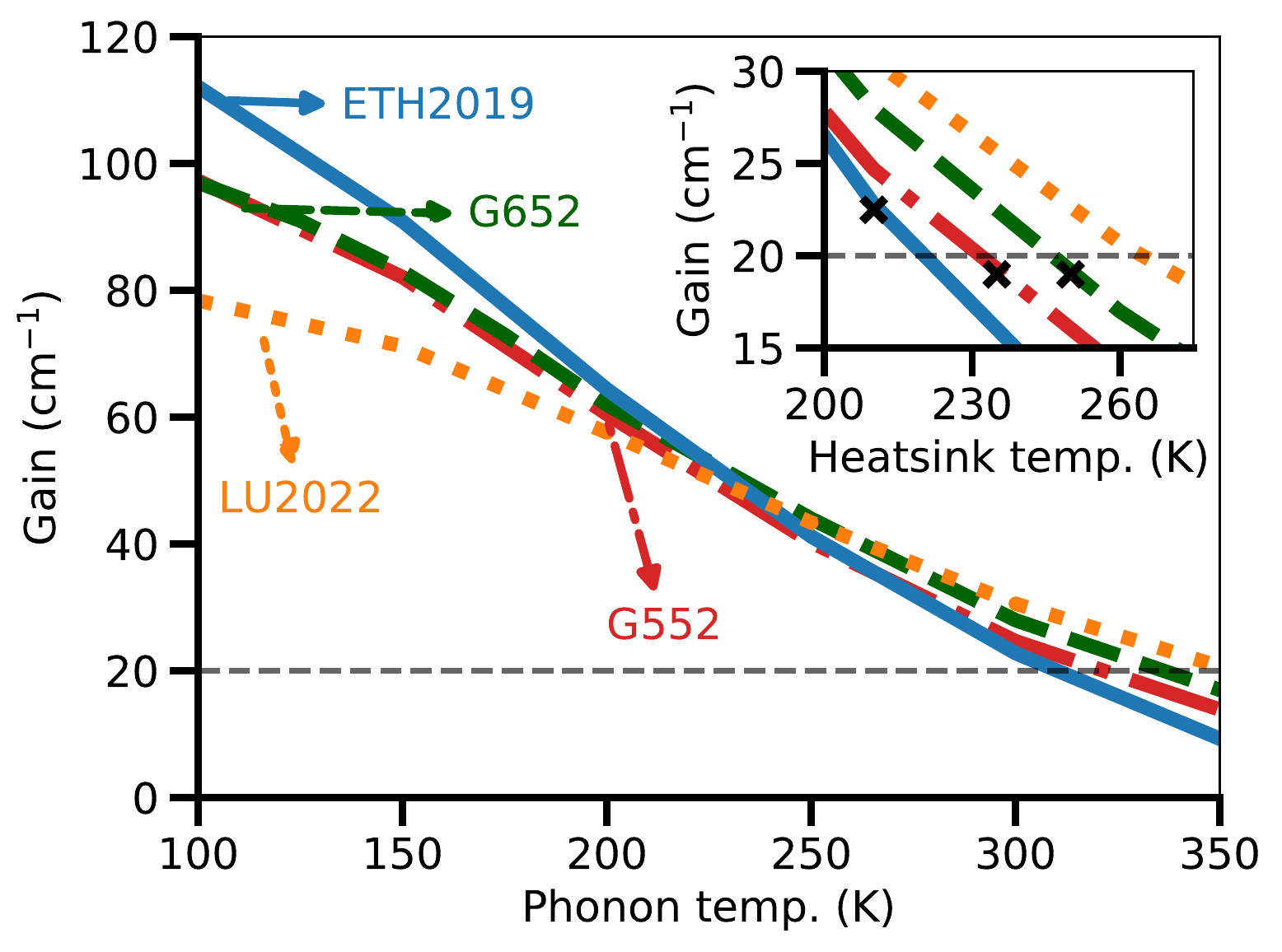} 
 \caption{Maximal gain as a function of phonon temperature for the
   samples from Refs.~\cite{BoscoAPL2019,KhalatpourNatPhot2021} and
   the design LU2022 suggested here. The peak gain was evaluated at
   the maximal current (without lasing) for the given temperature. The
   inset shows the same data for temperatures reduced by 90 K, which
   is our estimate for the difference between heatsink and phonon
   temperature. The crosses indicate the maximal heatsink temperature,
   where laser operation was observed  experimentally.}
 \label{fig:PeakGain}
\end{figure}

 In Fig.~\ref{fig:PeakGain}, we show the maximum gain (without lasing)
 obtained for different samples as a function of the phonon
 temperature.  We find that the device ETH2019 \cite{BoscoAPL2019}
 shows higher gain at low temperatures but gain is dropping stronger
 with temperature compared to the MIT devices
 \cite{KhalatpourNatPhot2021}. The condition
 that gain needs to compensate for losses allows us to determine the
 maximum phonon temperature for operation. In the inset of
 Fig.~\ref{fig:PeakGain} the crosses indicate the observed maximal
 heatsink operation temperatures. These agree well with the calculated
 maximal gain under the assumption, that the phonon temperature is 90
 K larger than the heatsink temperature reported in the experimental
 work. While we did not perform simulations on the phonon kinetics, we
 note that the order of magnitude of this shift is consistent with
 more detailed studies \cite{VitielloAPL2012,ShiJAP2014}. This holds
 in particular for the most relevant optical phonons, which are
 expected to become quickly excited even for short pulses. Thus our
 simulations provide a good description for the differences in maximal
 lasing temperature of the samples.

\begin{figure} 
	\includegraphics[scale = 0.25]{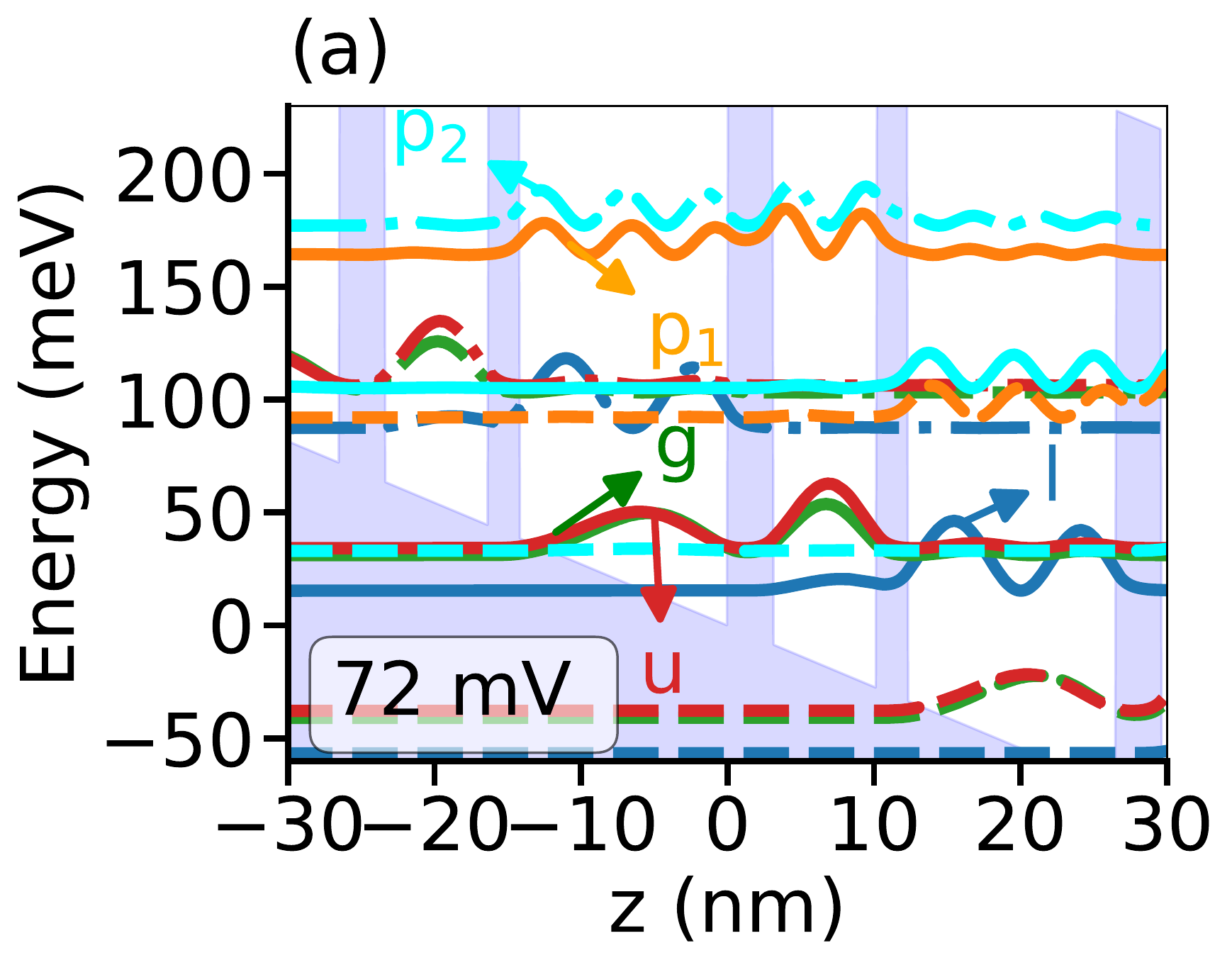}
        \includegraphics[scale = 0.25]{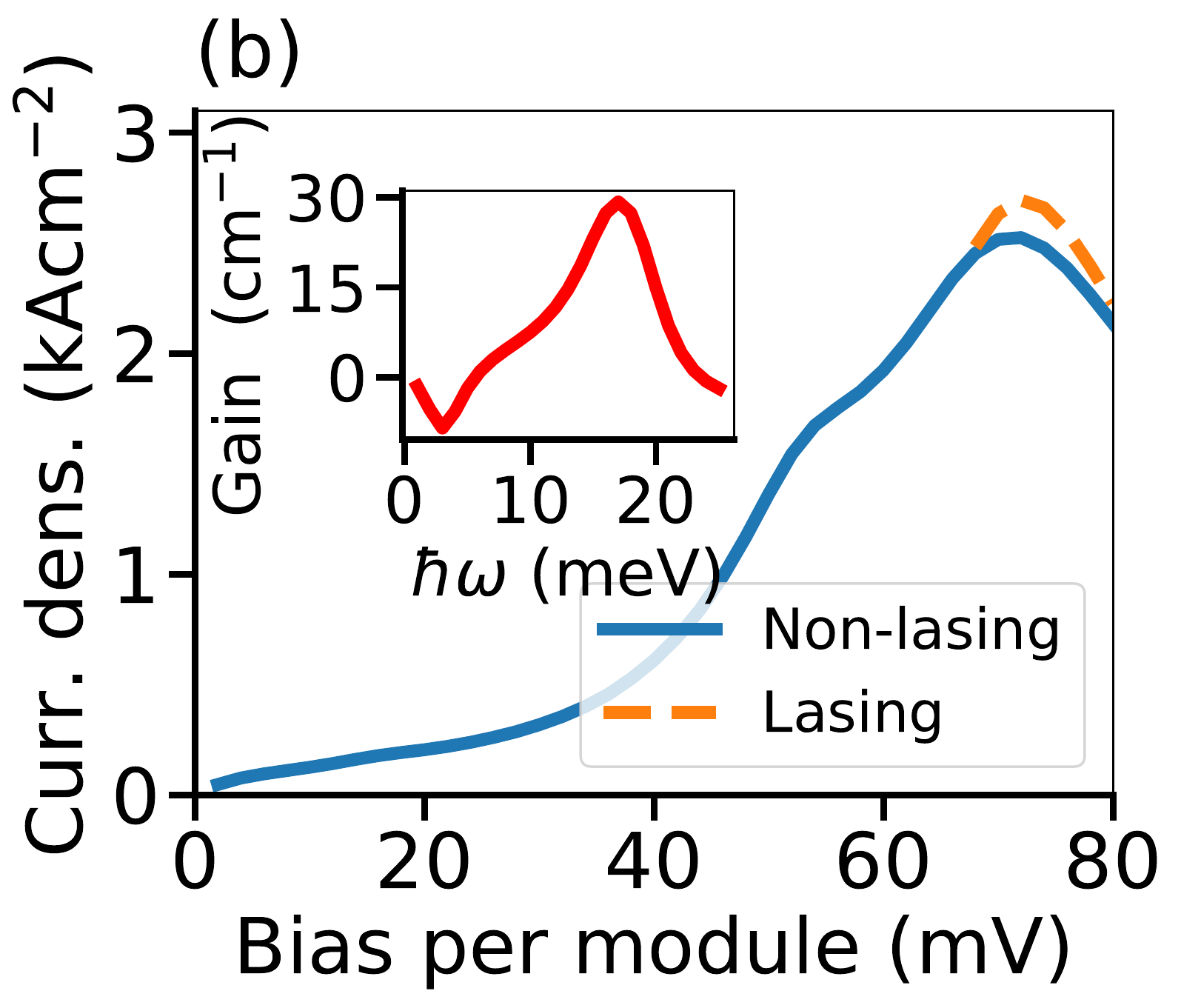}

 \caption{The proposed device LU2022 with a layer sequence ${\bf
     31}/71/{\bf 21}/142$ (in {\AA} with x=0.3 barriers in bold face),
   where the central 3 nm of the largest well are n-doped with
   $4.5\times 10^{10}\textrm{cm}^{-2}$ per module. Left: Band diagram
   at peak current. Right: Current bias relation for
   $T_\textrm{ph}=300$ K. The inset shows the gain spectrum at the
   current peak without lasing. }
 \label{fig:LU2022}
\end{figure}

In Fig.~\ref{fig:PeakGain} we added an additional design, called
LU2022, which shows even better temperature stability. From the
intersection of the peak gain with the expected losses of 20/cm in the
inset, we expect a maximal operation temperature around 265 K. Details
on this design are given in Fig.~\ref{fig:LU2022}. We note that the
current densities are comparable to the MIT devices and lower than the
ETH device, which demonstrates that the device is manageable in
existing setups. In these calculations  we used the band offset $\Delta
E^2_c$, as appropriate for MIT samples. Applying instead $\Delta
  E^1_c$, we get almost identical results for an increased Al
  concentration $x=0.35$. This indicates that an $x$-value slightly
  above 0.3 might be appropriate for other labs.
 
\begin{figure} 
		\centering
                \includegraphics[width=0.8\linewidth]{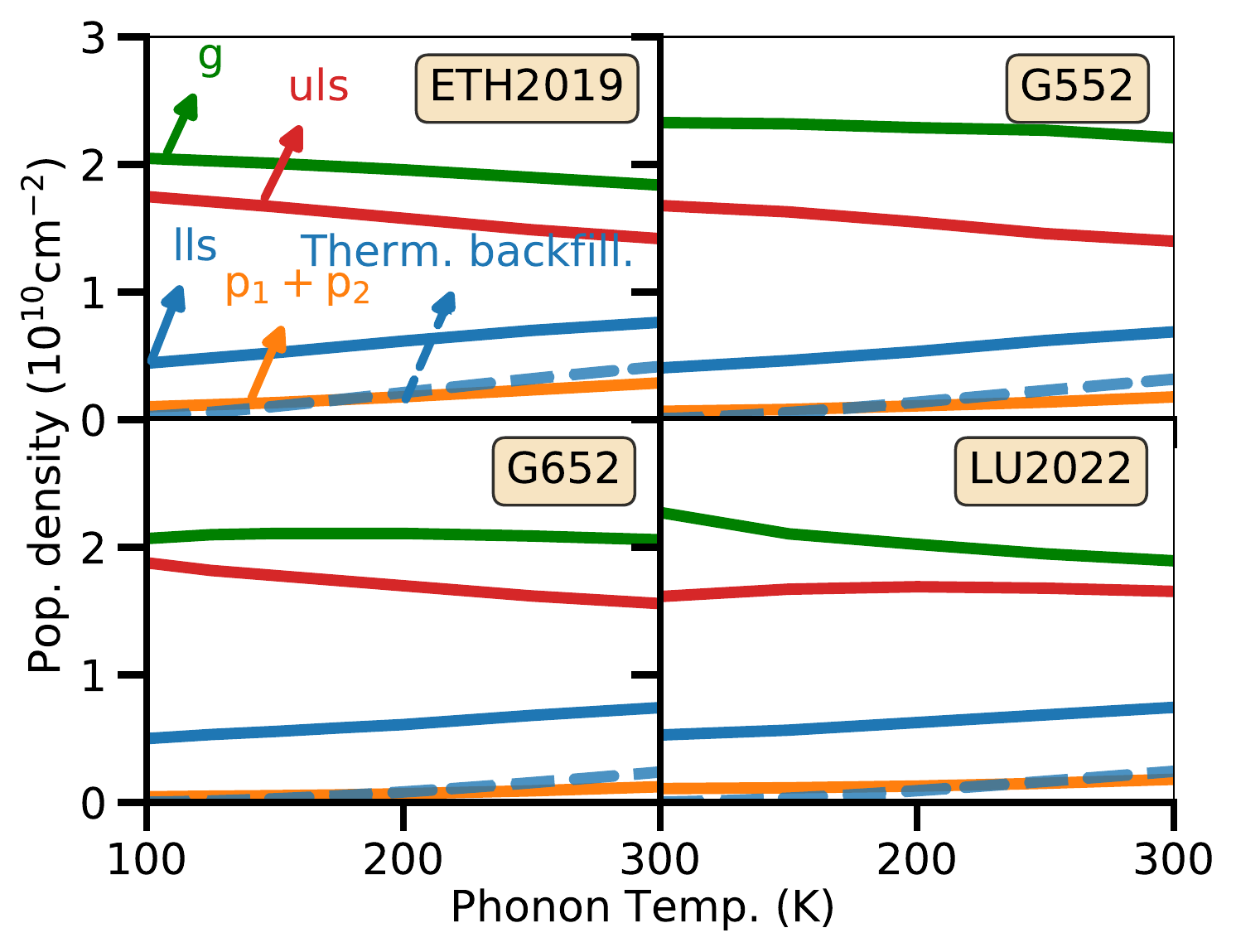}
 \caption{Calculated occupation of the energy eigenstates (full lines)
   and estimated contribution from thermal back-filling for the lower
   laser level  (dashed line) for different
   samples. }\label{fig:Backfilling}
\end{figure}

In order to analyze the temperature dependence, we  extracted the
occupations of the levels as a function of temperature for the bias
with maximal gain.  These are displayed in Fig.~\ref{fig:Backfilling},
where we added the expression $n_g\e^{-(E_l-E_g)/k_BT_\textrm{ph}}$
due to thermal back-filling from the ground level $g$. We find that
thermal back-filling provides the essential  ingredient in the thermal
increase in population of the lower laser level. Here the sample
ETH2019 has a level separation $E_l-E_g=38$ meV at the peak gain. This
approximately equals the longitudinal optical phonon energy and
provides the quickest emptying for the lower laser level
\cite{FerreiraPRB1989} and is traditionally used in QCL designs. In
contrast the samples MITG552, MITG652, LU2022 exhibit a significantly
larger separation of 47 meV, 53 meV, and 53 meV, respectively, which
reduces thermal back-filling. We suggest that this can relate to the
better temperature stability visible in Fig.~\ref{fig:PeakGain}. This
had been already addressed in \cite{FranckiePhysRevApplied2020} for
two-well designs and is actually the same scenario as observed for
infrared QCLs, where the reduction of thermal back-filling due to
double phonon extraction allowed for room temperature operation
\cite{BeckScience2002}. On the other hand, ETH2019 has the highest
gain at low phonon temperatures, showing that it is actually the
better design here.

\begin{figure} 
	\includegraphics[scale = 0.223]{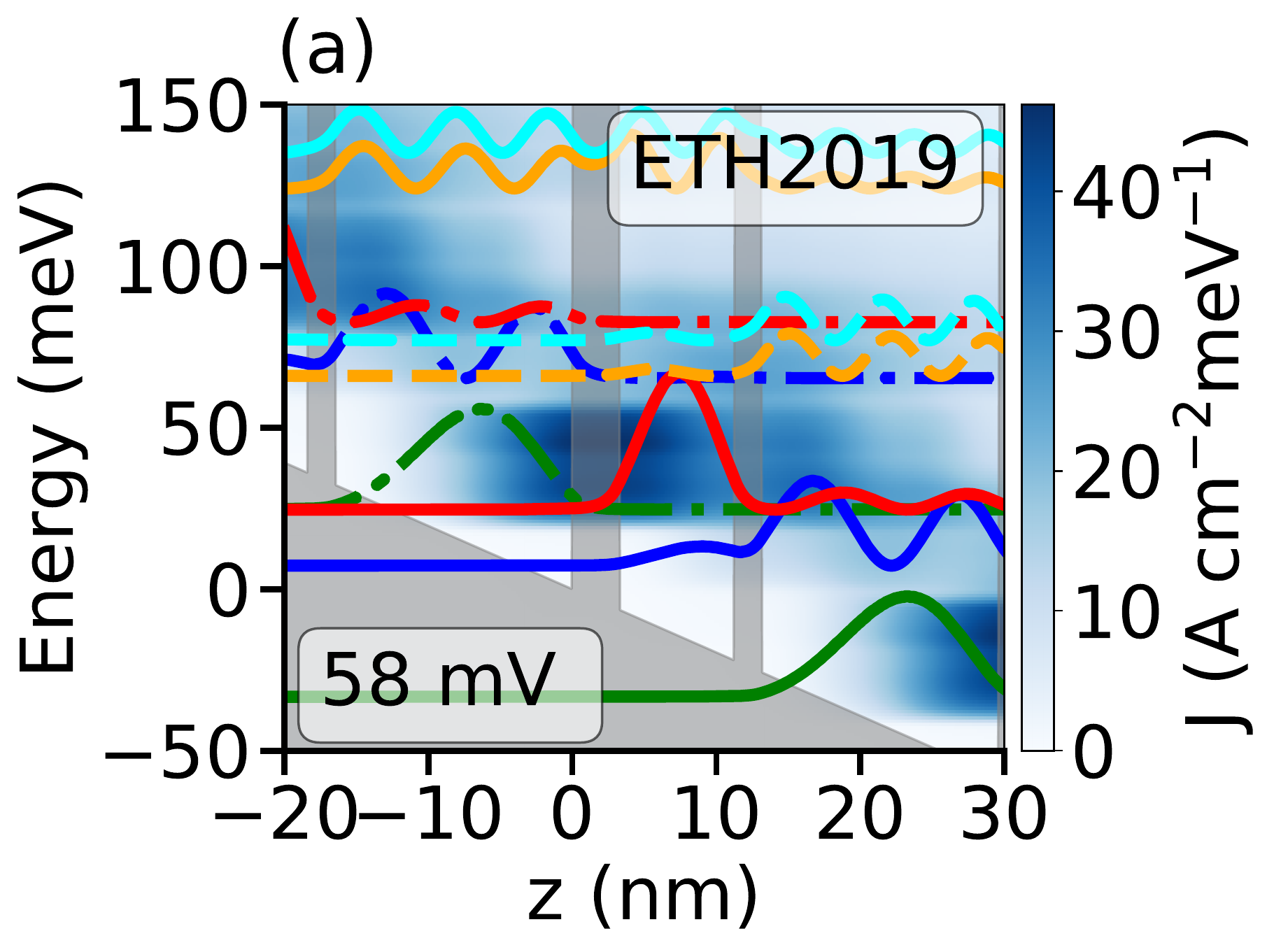}
        \includegraphics[scale = 0.223]{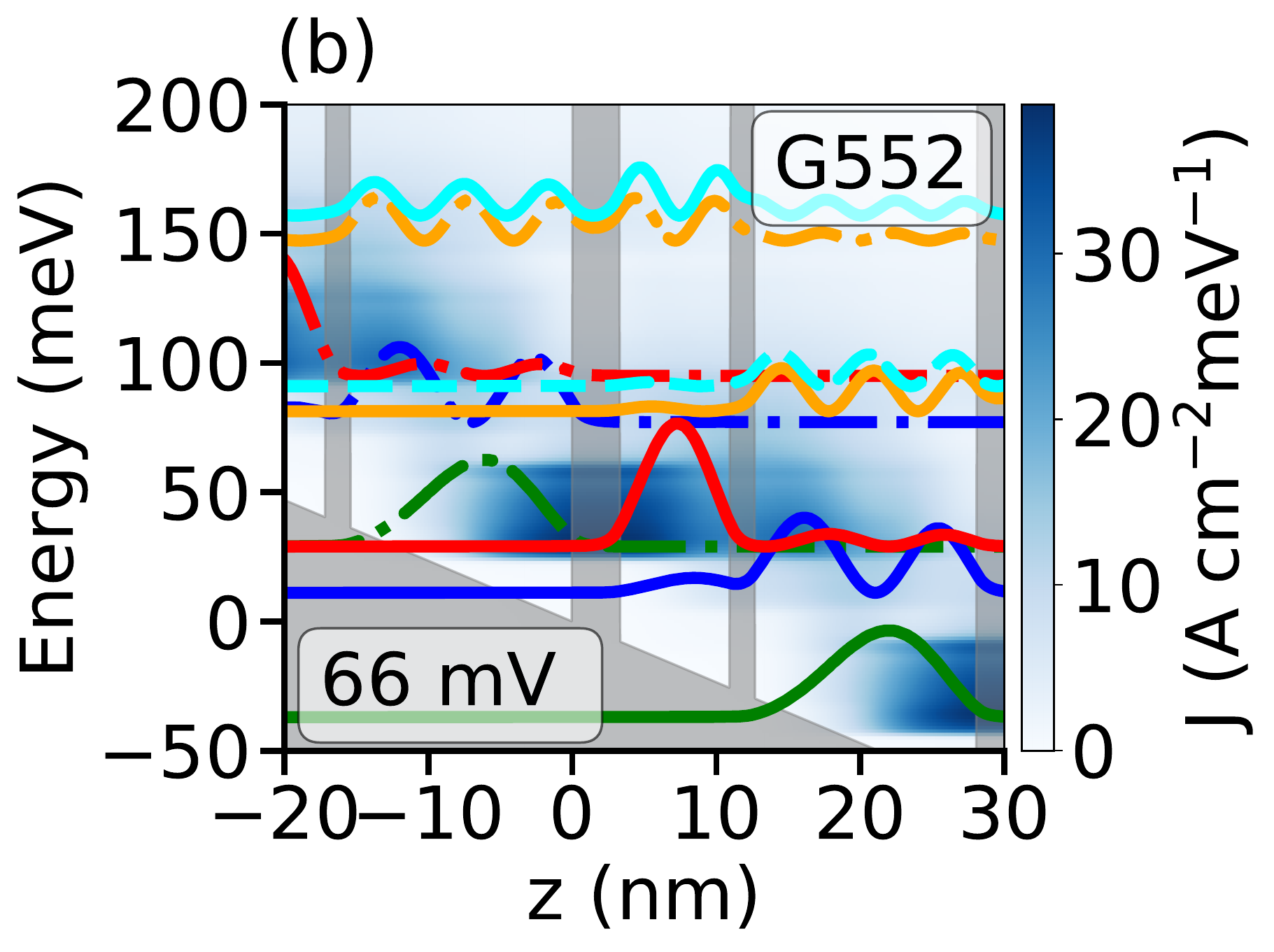}
        \includegraphics[scale = 0.223]{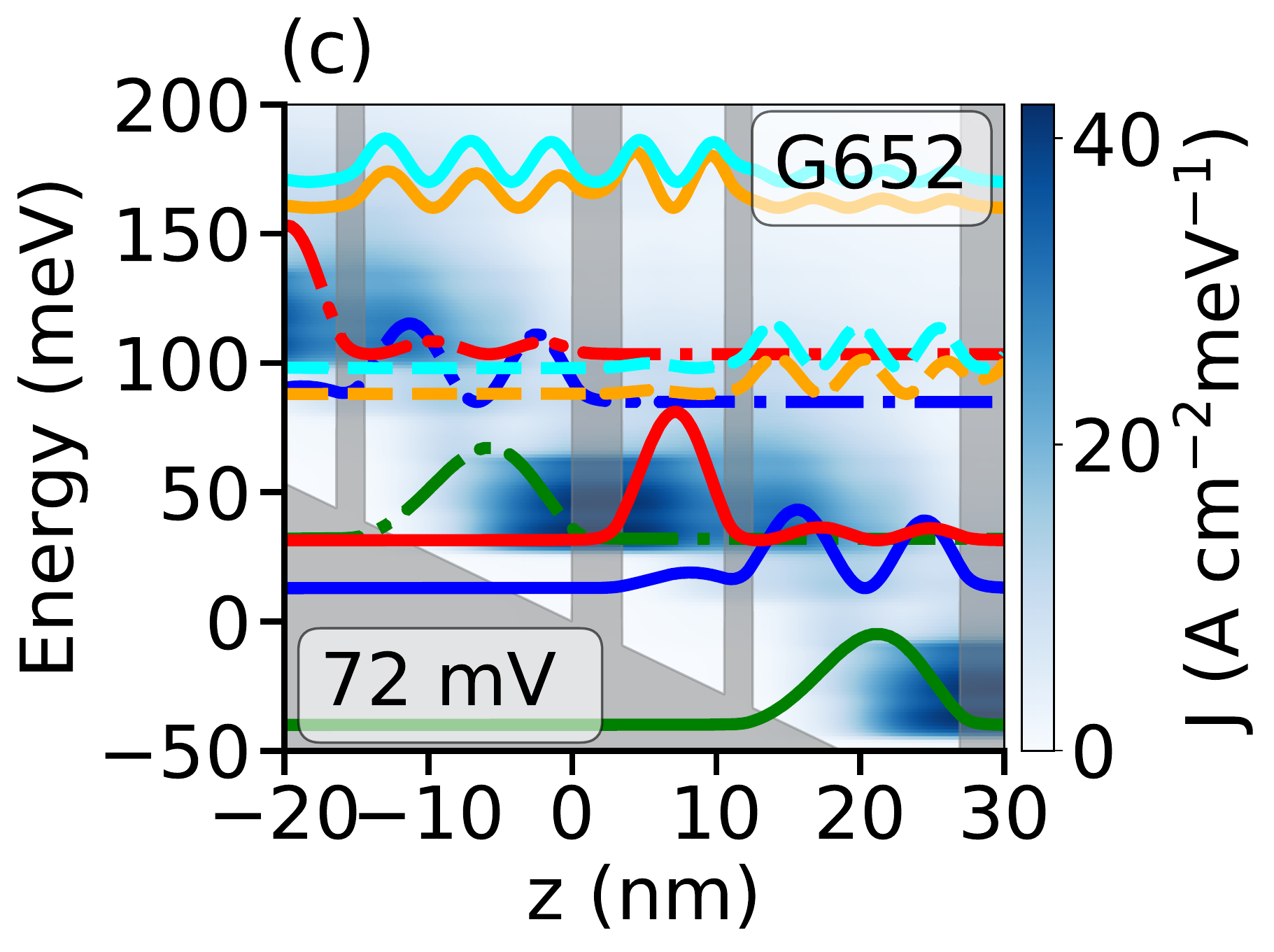}
        \includegraphics[scale = 0.223]{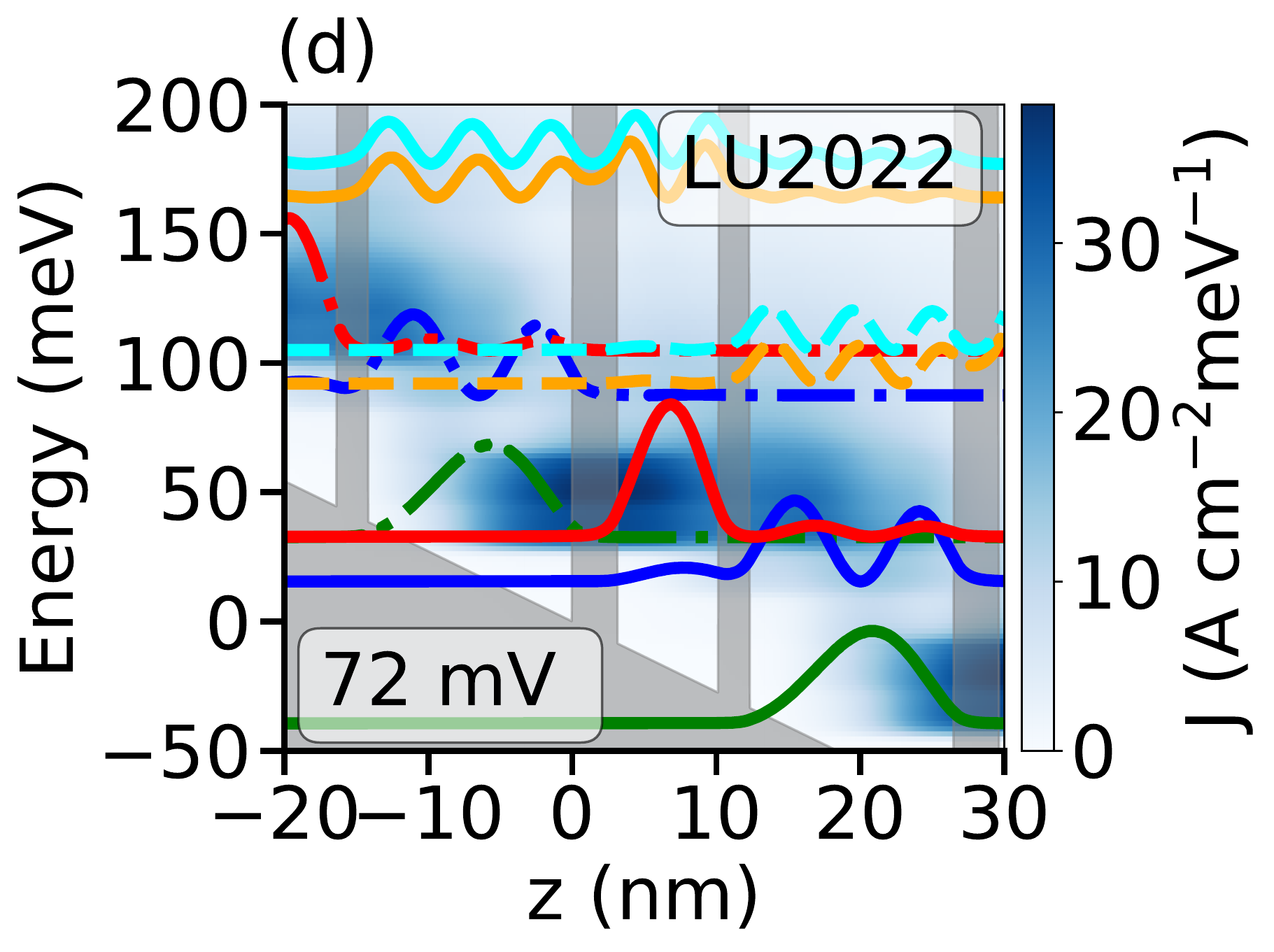}
 \caption{Calculated current distribution for different samples at
   peak bias and a phonon temperature of 350 K together with the EZ
   states defined in the text below. }\label{FigResolveCurrent}
\end{figure}

Ref.~\cite{KhalatpourNatPhot2021} stated that a pure 3 level design
should be favored as this prevents from electrons
escaping. Fig.~\ref{fig:Backfilling} shows that the ETH sample has
indeed a larger occupation of the parasitic levels (denoted as p1 and
p2 here) at high temperature compared to the top performing devices
MITG652 and LU2022.  This can also be seen in the spatio-energetically
resolved current densities in Fig.~\ref{FigResolveCurrent}, where
higher states contribute more for the ETH sample. In all cases the
levels p1 and p2 are energetically close to the lower and upper laser
level. Thus the key mechanism is the use of higher barriers reducing
resonant currents and the larger vertical energy separation, reducing
excitations from the u, l, g levels into the parasitic states.

The decrease of gain with temperature
(see Fig.~\ref{fig:PeakGain}) is much stronger than one would expect
based on the decrease in inversion $n_u-n_l$ as calculated by the
common energy eigenstates (see Fig.~\ref{fig:Backfilling}).  In
Ref.~\cite{NelanderAPL2008} such a difference was related to the
scattering rates increasing with temperature, which result in a
stronger broadening of the gain spectrum. For the data presented here,
we also see this trend, but it is by far not strong enough to  provide
the substantial decrease in gain.

\begin{figure} 
\includegraphics[width=\columnwidth]{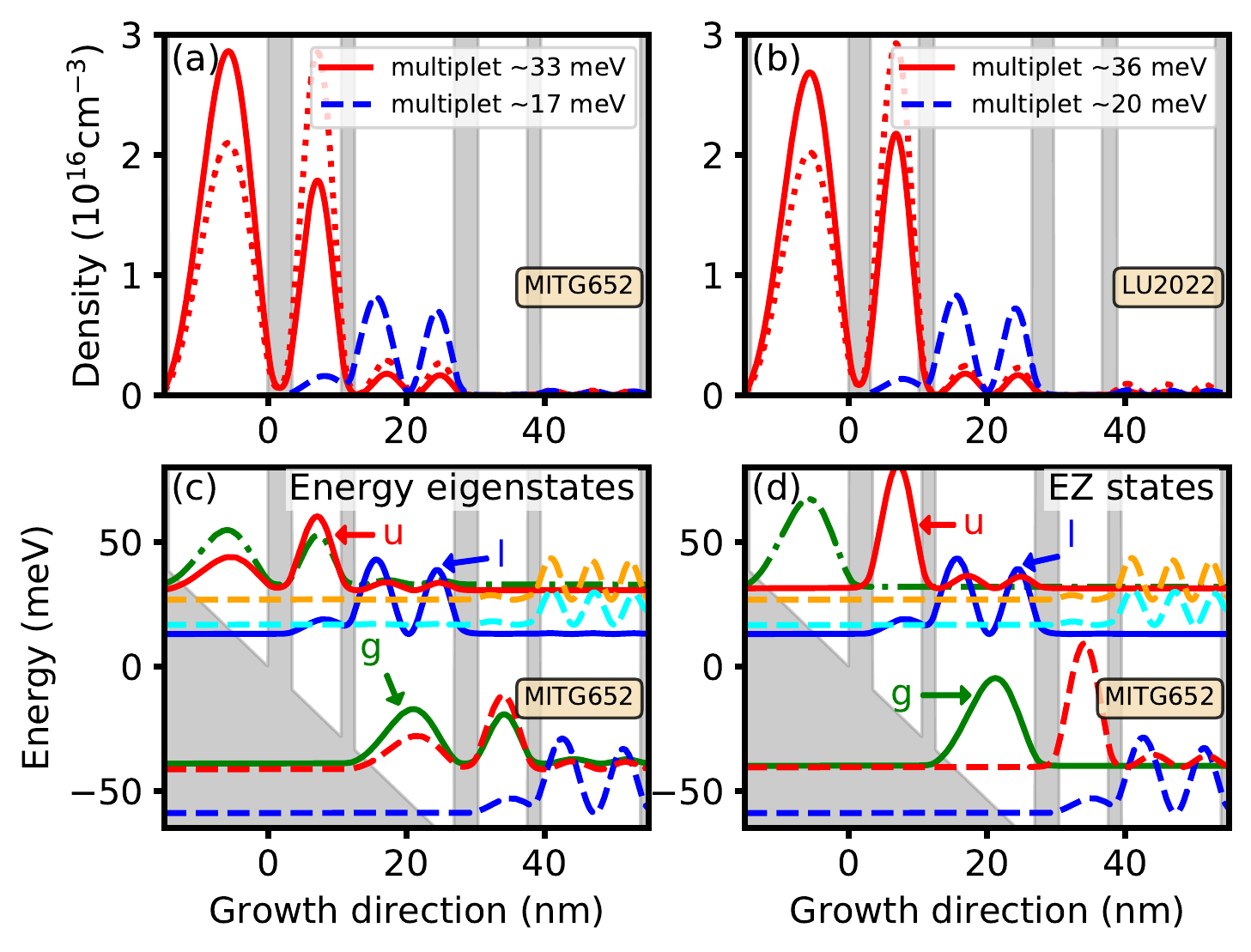}
\caption{Upper panels: Electron density from Eq.~\eqref{eq:density}
  for the multiplet including the upper laser level (red full) and the
  multiplet including the lower laser level (blue dashed) for the
  sample MITG652 (a) and LU2022 (b). The red dotted line restricts to  the diagonal elements of $\rho_{ij}$ for the upper level multiplet.  Lower panels: Absolute square of
  wave functions for the sample MITG652.  (c) Energy eigenstates; (d)
  EZ states. All data is extracted from our NEGF simulations at
  $T_\textrm{ph}=350$ K at the peak bias of 72 mV/module.}
\label{fig:CompareEZ_WS}
\end{figure}

It is well known \cite{CallebautJAP2005,WackerPRL1998} that the energy
eigenstates as e.g. shown in Figs.~\ref{fig:LU2022}(a) and \ref{fig:CompareEZ_WS}(c)
are not a good
basis if the energy separation is smaller than the broadening. In this
case the coherences between the states  must be carefully taken into
account. Here we achieve this by identifying multiplets of energy
eigenstates $|\Psi_j\rangle$, which are close in
energy. (Specifically, we add a state to a multiplet if its energy
differs by less than 5 meV to at least one other member of the
multiplet.)
Restricting the general expression for the spatial electron density $n(z)$ to the states of the multiplet, we get
\begin{equation}
n_\textrm{multiplet}(z)=\sum_{i,j\in \textrm{multiplet}}
\rho_{ij}\Psi^*_j(z)\Psi_i(z)\label{eq:density}
\end{equation}
where $\Psi_i(z)$ are the wavefunctions of the states. 
$\rho_{ij}$  is the density matrix within this basis, where the diagonal elements provide the occupations of the states (in 1/cm$^2$) and the nondiagonal elements (also called polarizations) shift the electron density in space. Results are shown in
the upper panels of Fig.~\ref{fig:CompareEZ_WS} for the devices
MITG652 (a) and LU2022 (b).  In both cases, the electrons mostly stay
on the left side of the injection barrier around $z=0$ (red full line).
This is not well reflected by the sum of densities of the energy eigenstates (red dotted line), where the occupations of the upper laser
state and the ground state are pretty similar at
resonance, see Fig.~\ref{fig:Backfilling}. Furthermore, we note, that the injection to the right side
of the barrier, from where optical transitions to the lower laser
level occur, is better for LU2022, which has a slightly thinner
injection barrier.  This appears the key point for the better
performance of LU2022 at high temperatures.

In order to quantify the injection, is is better to use a localized
basis. For this purpose we calculate the matrix
$\langle\Psi_i|\hat{z}|\Psi_j\rangle$ for each
multiplet. Diagonalizing this matrix provides the basis transformation
from the energy eigenstates of each multiplet to localized states,
which we call EZ-states in the following.  (The name is motivated by
the reasonably well defined energy $E$ and position $z$ within the
subspace of the multiplet). In the lower panels of
Fig.~\ref{fig:CompareEZ_WS}, we show a comparison between the energy
eigenstates (c) and the EZ states (d), showing that the ground (green)
and upper laser level (red) are much better defined for the EZ
states. In contrast, the corresponding energy eigenstates represent
binding and anti-binding combinations, where the absolute square of
the wave function is very similar at resonance. By construction these
EZ states provide an orthonormal basis, where the diagonal elements of
the Hamiltonian (transformed into this basis) provide the energy of
the states and nondiagonal elements represent tunnel couplings
$\Omega$ between almost degenerate states within a multiplet.

\begin{figure} 
		\centering
                \includegraphics[width=0.8\linewidth]{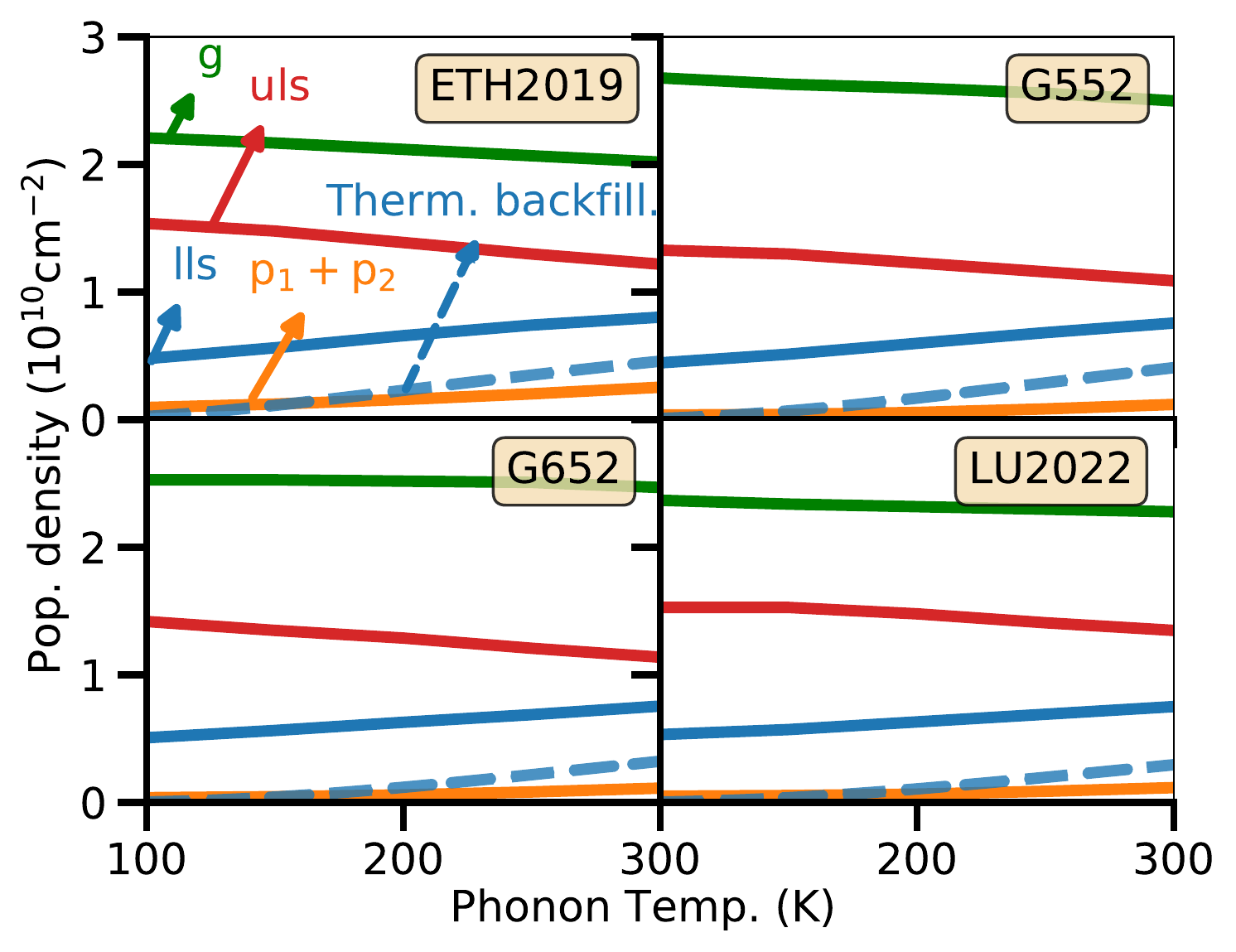}
 \caption{Calculated occupations of the EZ states (full lines) and
   estimated contribution from thermal back-filling for the lower laser
   level  (dashed line) for different samples.
 }\label{fig:EZoccupations}
\end{figure}

Fig.~\ref{fig:EZoccupations} shows the occupations of the EZ states
(obtained from the diagonal elements of the density matrix transformed
to EZ states). Here the occupations of the upper laser level are
significantly lower than the occupations of the ground
level. Furthermore, the difference between the upper and lower laser
levels decreases drastically with temperature, explaining the major
part of the temperature drop in gain.  

Considering the EZ states in Fig.~\ref{fig:EZoccupations}, the
difference between the ground and upper laser level hardly changes
with temperature for all devices. This difference is actually
proportional to the injection current in the device, which has no
strong temperature dependence at the current peak, see also
\cite{FreemanJAP2020}. For sequential tunneling, where the broadening
$\Gamma$ surpasses the tunnel coupling $\Omega$, we find similar to
Eq.~(83) of \cite{WackerPhysRep2002}
\begin{equation}
J_{i\to u}^\textrm{seq. tunnel.}=\frac{e\Omega^2}{\hbar}
\frac{\Gamma}{(E_u-E_i)^2+\Gamma^2/4} (n_i-n_u)
\end{equation}
which provides very good agreement for all samples with
$\Gamma\sim 6-8$ meV.
Here we extracted $\Omega=1.58,\, 1.08,\,  1.17,\textrm{ and
}1.49$ meV for the samples ETH2019, MITG552, MITG652, and LU2022,
respectively, from the Hamiltonian transferred to the EZ basis. (The
values for the MIT samples agree very well with the anticrossing gaps
$\Omega_{iu}$ reported in Table 1 of \cite{KhalatpourNatPhot2021},
which correspond to $2\Omega$ in our notation.) The rather low value
of $\Omega$ explains the large differences between $n_u$ and $n_i$ for
the MIT samples. This actually limits the inversion, as it provides an
upper bound for $n_u$. In contrast our design LU2022 has a thinner
injection barrier, while it restricts the current by a larger lasing
barrier, which results in better behavior.  On the other hand, for
ETH2019 the injection in more efficient (even if it is taken into
account that the currents are higher). Here the limiting factor is the
higher thermal back-filling and the population of enhanced higher
states addressed above. 

\textit{Conclusion:} Using NEGF simulations, we could quantitatively reproduce the observed
maximal operation temperatures of high-performing three-level
designs. We confirm that the occupation of higher states with
increasing temperature and thermal back-filling is detrimental for the
design of \cite{BoscoAPL2019}. Both issues are improved in the MIT
designs \cite{KhalatpourNatPhot2021} using higher barriers. However
these MIT designs have a rather thick injector barrier leading to a
significant difference between the occupations of the ground and upper
laser state, which limits inversion. This issue can be conveniently
analyzed by using EZ states. These are obtained from the energy
eigenstates (E) which are sorted to multiplets with similar
energy. Then the $Z$-operator is diagonalized within each multiplet
subspace, providing the EZ states.  The analysis shows, that the
devices of Ref.~\cite{KhalatpourNatPhot2021} suffer from insufficient
injection. Using a thinner injection barrier while limiting the
current increase by a thicker radiation barrier, we could identify a
design which is estimated to operate up to 265 Kelvin.

\textit{Acknowledgments:} We thank the Swedish Research Council (project
2017-04287) and NanoLund for financial support.

%

\end{document}